\documentstyle[amssymb,12pt]{article}
\begin{document}
\title{ON CONSTRUCTION OF RECURSION OPERATORS FROM LAX REPRESENTATION}
\author{ Metin G{\" u}rses\\
{\small Department of Mathematics, Faculty of Sciences}\\
{\small Bilkent University, 06533 Ankara - Turkey}\\
Atalay Karasu\\
{\small Department of Physics, Faculty of Arts and  Sciences}\\
{\small Middle East Technical University, 06531 Ankara-Turkey}\\
Vladimir V. Sokolov\\
{\small Landau Institute,  Moscow , 117940 Russia}}
\date{}
\begin{titlepage}
\maketitle

\begin{abstract}
In this work we develop a general procedure for constructing the
recursion operators for non-linear integrable equations admitting
Lax representation. Several new examples
are given. In particular we find the recursion operators for
some KdV-type systems of integrable equations.

\end{abstract}
\end{titlepage}

\section{Introduction}

It is well known that most of the integrable nonlinear partial differential
equations

\begin{equation}
u_{t}=F(t,x,u,u_{x}, \cdots,u_{nx}),  \label{b0}
\end{equation}

\noindent
admit Lax representation

\begin{equation}
L_{t}=[A,L] , \label{l0}
\end{equation}

\noindent
so that inverse scattering method is applicable. The generalized
symmetries \cite{olv} of (\ref{b0}) have also Lax representations
with the same $L$-operator

\begin{equation}
L_{t_{n}}=[A_{n},L]~,~~n \ge 1. \label{l2}
\end{equation}

\noindent
The recursion operator $\cal{R}$, satisfying the equation (see \cite{olv1})

\begin{equation}
{\cal{R}}_{t}+[D_{F},{\cal{R}}]=0, \label{b1}
\end{equation}

\noindent
where $D_{F}$ is the Frech{\' e}t derivative of the function $F$,
generates symmetries of (\ref{b0}) starting
from the simplest ones. In general $\cal{R}$ is a nonlocal operator (a pseudo-differential
operator).

The construction of the recursion operator of a given integrable
system (\ref{b0})
is not an easy task. Several works are devoted to this subject. Among these
works most of the authors use (\ref{b1})
for the construction of the recursion operator \cite{gur1}-\cite{kra}.
There are several difficulties in this direct approach.
The main problems are the choices of the
order of ${\cal{R}}$ and the structure of the nonlocal terms. This is an
approach, having no relation with the Lax representation (\ref{l0}).

On the other hand some of the authors used Lax representation
for this purpose.
Most of these works are related to the squared
eigenfunctions of the Lax operator \cite{fok0}- \cite{san2}
and are based on finding eigenvalue equation
for the squared eigenfunctions of the Lax operator.
The operator corresponding to this eigenvalue equation
turns out to be the adjoint of the recursion operator.

There is an alternative use of the Lax representation to construct
recursion operators. This approach is based on the explicit
construction of the
$A_{n}$- operators (\ref{l2}). It was first used by Symes \cite{sym}, Adler
\cite{adl} (see also Dorfman-Fokas \cite{dor}, Fokas-Gel'fand \cite{fok5})
and Antonowicz-Fordy \cite{for2},\cite{for3}. Although these authors
use the Lax representation in different ways their approach is basically
the same. Symes and Adler use the Gel'fand-Dikii \cite{gd} construction of
the $A_{n}$- operators.
On the other hand Antonowicz-Fordy determines these operators from
integrability condition (\ref{l2})
and  by using an ansatz for $A_{n}$. Their basic aim is to
determine the Hamiltonian operators $\theta_{1}$ and $\theta_{2}$ \cite{mag}
of the equations under consideration. The recursion operator is simply
given by ${\cal R}=\theta_{2}\, \theta_{1}^{-1}$. Their approach is based on
some explicit formulas for coefficients of $A_{n}$-operator. This is
necessary to find the Hamiltonian operators $\theta_{1}$ and $\theta_{2}$
and it seems that this approach is quite effective to determine the
bi-Hamiltonian structure for the simple cases but it becomes more
complicated when $L$-operator has a sophisticated structure.

If one is interested only in the determination of the recursion operator
$\cal{R}$, we shall show in this work that it is possible to succeed this
without any concrete information of the coefficients of $A_{n}$-operators.
We use only an ansatz $\tilde{A}={\cal{P}}\,A+R$ that
relates $A_{n}$-operators for different $n$. Here $\cal{P}$ is some
operator which commutes with the $L$-operator and $R$ is the remainder.

We follow this basic idea, partially used by Symes \cite{sym}, Adler
\cite{adl}, Shabat and Sokolov \footnote{In 1980 Shabat and Sokolov
independently found the recursion operator for the Sawada-Kotera
equation. This result was published in \cite{ibr}. In \cite{sok0}
Sokolov found the recursion operator for the Krichever-Novikov equation.}
\cite{sok0}, and establish an extremely simple, effective and algorithmic
method for the construction of recursion operators when the Lax representation
(\ref{l0}) is given.

In the next section we consider the case where $L$ is a
scalar operator. We first consider the case where $L$ is a
differential operator and then the case where it is a
pseudo-differential operator.
In each case we present our method, discuss the reductions
and give examples for illustrations.
In section 3 we consider Lax operator taking values in a Lie algebra.
We give our method both for the general case and also for the reductions.
We give one example for each
case in the text. Several additional examples are given in the Appendices
A, B, and C corresponding to all different cases.

\section{Scalar Lax representations.}

First we consider equations with the scalar Lax representations of the
form

\begin{equation}
 L_{t}=[A, \ L],     \label{a0}
\end{equation}

\noindent
where $L$ is, in general, a pseudo-differential
operator of order $m$ and $A$ is a differential operator
whose coefficients are functions of $x$ and $t$.

The different choice of operators $A$ for a given $L$ leads
to a hierarchy of nonlinear systems (\ref{l2}).
It is well known that one can define operators $A_n$ by the
following formula \cite{gd}

\begin{equation}
 A_{n}=(L^{n/m})_{+}, \label{a2}
\end{equation}

\noindent
where $L^{n/m}$ is a pseudo-differential series of the form
$L^{n/m}=\sum^n_{-\infty} v_i D^i$ and
$(L^{n/m})_+=\sum^n_{i=0} v_i D^i$. Here $v_{i}$ are some concrete
functions depending on the coefficients of $L$ and $D$ is the total
derivative with respect to $x$.

In  \cite{ds1},\cite{ds2} the relationships between the Kac-Moody
algebras and special types of scalar differential and pseudo-differential
operators $L$ were established. All corresponding integrable systems are
Hamiltonian ones. For most of them a second Hamiltonian structure is not
known up to now.

In this section and Appendices A,B, and C
we consider the simplest systems from \cite{ds1} and \cite{ds2} as examples
 and find their recursion operators. In the sequel these examples will
be referred as Drinfeld-Sokolov (DS) systems. It is interesting
to note that in all these examples the order of the recursion operator
is equal to the Coexter number of the corresponding Kac-Moody algebra.

\subsection{Gel'fand-Dikii Systems.}

In this section we shall consider the case where $L$ is a differential
operator,

\begin{equation}
L=D^m+u_{m-2} D^{m-2}+\cdots +u_0,
\end{equation}

\noindent
where $u_{i}, i=0,1, ..., m-2$ are functions of $x,t$.
In the framework of
\cite{ds1} this corresponds to the Kac-Moody algebras of the type
$A^{(1)}_{m-1}$.

 To show that (\ref{l2}) is equivalent to a system of $(m-1)$
evolution equations with respect to $u_{i}$ one can use the
following standard reasoning. Set

\begin{equation}
L^{n \over m}=(L^{n \over m})_{+}+(L^{n \over m})_{-}, \nonumber
\end{equation}

\noindent
where $(L^{n \over m})_{+}$ is the differential part of the series
$L^{n \over m}$ and $(L^{n \over m})_{-}$ is a series of order $ \le -1$.
Since $[L, L^{n \over m}]=0$ we have

\begin{equation}
[(L^{n \over m})_{+},~~ L]=[L,~~(L^{n \over m})_{-}]. \label{gd1}
\end{equation}

\noindent
The left hand side of (\ref{gd1}) is a differential operator,
but the right side is a series of order $\le n-2$. Thus both side
of (\ref{l2}) are differential operators of order $\le n-2$ and
it is equivalent to a system of evolution equations for
the dependent variables $u_{i}, i=0,1, ..., m-2$. This system can be
obtained by comparing the coefficients of $D^{i}$ where $0, \cdots m-2$
in (\ref{l2}).

Since $L^{n+m \over m}=L\, L^{n \over m}$ then we have

\begin{equation}
A_{m+n}=(L\, L^{n \over m})_{+}=L\,( L^{n \over m})_{+}+
(L \, (L^{n \over m})_{-})_{+}, \label{l00}
\end{equation}

\noindent
which leads directly to

\begin{equation}
 L_{t_{n+m}}=[A_{n+m}, \ L]= L L_{t_{n}}+[(L\,(L^{n \over m})_{-})_{+},\, L].
 \label{l01}
\end{equation}


\noindent
The above equation (\ref{l01}) has been given also by Symes \cite{sym}
(see also Adler's paper \cite{adl}). In his work
Symes expressed  the coefficients of the both parts of (\ref{l01}), in a
rather complicated way, in terms of some finite set of coefficients of the
resolvent for $L$- operator. That allows him to express $L_{t_{n+m}}$
in terms of  $L_{t_{n}}$. This relation gives directly the recursion
operator. He gave explicit formulas for the cases $m=2$ and $m=3$.

In this section we shall show that in order to construct the
recursion operator it suffices to know only that

\begin{equation}
L_{t_{n+m}}=L\, L_{t_{n}}+[R_{n}, L]. \label{e00}
\end{equation}

\noindent
Obviously, it follows from the following

\vspace{0.3cm}

\noindent
{\bf Proposition 1}. For any $n$

\begin{equation}
A_{n+m}=L A_n+R_n, \label{aa}
\end{equation}

\noindent
where $R_{n} $ is a differential operator of order $\le m-1$.

\noindent
{\bf Proof:}\,\, The relation  (\ref{aa}) coincides with (\ref{l00})
if we put

\begin{equation}
R_{n}=(L(L^{n \over m})_{-})_{+}.
\end{equation}

\noindent
Since $(L^{n \over m})_{-}$ is a series of order $\le -1$, then
$ord\, (R_{n}) \le m-1$.

\vspace{0.3cm}

\noindent
{\bf Remark 1:}\, It follows from the formula

\begin{equation}
A_{n+m}=(L^{n \over m}\,L)_{+}=(L^{n \over m})_{+}\,L+
((L^{ n \over m})_{-}\,L)_{+},
\end{equation}

\noindent
that

\begin{equation}
A_{n+m}=A_{n}\,L+\bar{R_{n}}, \label{x00}
\end{equation}

\noindent
and

\begin{equation}
L_{t_{n+m}}=L_{t_{n}}\,L+[L, \bar{R_{n}}], \label{e01}
\end{equation}

\noindent
where $\bar{R_{n}}$ is a differential operator of order $\le m-1$.

To find the recursion operator we can simply equate the coefficients of
different powers of $D$ in (\ref{e00}). It is easy to see that in this
comparison of the coefficients of $D^{i}, \, i=2m-2,...,m-1$
we determine $R_{n}$ in terms of the coefficients of operators
$L$ and $L_{t_{n}}$. It is important that the resulting formulas
turn out to be linear in
the coefficients of $L_{t_{n}}$.
The remaining coefficients of $D^{i}, \, i=m-2,\cdots ,0$ in (\ref{e00})
give us the relation

\begin{equation}
\left(\matrix{u_{0} \cr
.\cr.\cr.\cr u_{m-2}} \right)_{t_{n+m}}={\cal R}\,
\left(\matrix{u_{0}\cr.\cr.\cr.\cr  u_{m-2}} \right)_{t_{n}} , \nonumber
\end{equation}

\noindent
where $\cal R$ is a recursion operator. Instead of (\ref{e00})
one can use (\ref{e01}). The corresponding recursion operators
coincide.

\vspace{0.3cm}

\noindent
{\bf Example 1. KdV Equation}. \, The KdV equation

\begin{equation}
 u_{t}={1 \over 4}(u_{3x}+6uu_{x}),     
\end{equation}
has  a  Lax representation  with

\begin{equation}
 L=D^{2}+u, \qquad A=(L^{3/2})_{+}.
\end{equation}

\noindent
Since in this case $L_{t_{n+2}}=u_{t_{n+2}} \equiv u_{n+2}$ and
$L_{t_{n}}=u_{t_{n}} \equiv u_{n}$ the main relation (\ref{e00})
takes the form

\begin{equation}
u_{n+2}=(D^2+u)\cdot u_{n}+[R_{n},L],
\end{equation}
with $R_{n}=a_{n}D+b_{n}$.

Now if we equate successively to zero the coefficients of $D^2$, $D$
and $D^0$ in above equation, we obtain

$$a_{n}={1 \over 2}D^{-1}(u_{n}),~~~ b_{n}={3 \over 4}u_{n}, $$

\noindent
and

$$u_{n+2}=({1 \over 4}D^{2} +u+{1 \over 2}u_{x}D^{-1})u_{n}, $$

\noindent
that gives the standard recursion operator for KdV equation

\begin{equation}
{\cal{R}}= {1 \over 4}D^{2} +u+{1 \over 2}u_{x}D^{-1}.
\end{equation}

In the same way one can find a recursion operator for the
Boussinesq equation (see Appendix A).

\subsection{Symmetric and skew-symmetric reductions\\ of differential Lax
operator.}

The standard reductions of the Gel'fand-Dikii systems are given by
the conditions $L^{*}=L$ or $L^{*}=-L$. Here $*$
denotes the adjoint operation defined as follows. Let $L$ be a
differential operator,
 $L=\Sigma \,a_{i}\,D^{i}$. Its adjoint $L^{*}$ is given by
$L^{*}=\Sigma\, (-D)^{i}\, \cdot a_{i}$. It is easy to see
that if $L^{*}=L$ then $m= ord\,(L)$ must be an even integer.
For the case $L^{*}=-L$ it must be an odd integer.

It is well known that for both reductions all
possible $A_{n}$ are defined by (\ref{a2}), where $n$ takes odd integer values.
This condition provides that $(A_{n})^{*}=-A_{n}$ which is necessary
for (\ref{l2}) to be compatible.

If $L^{*}=L$, the formula
$A_{n+m}=(L\,L^{n \over m})_{+}=(L^{n+m \over m})_{+}$
gives a correct $A_{n}$-operator since $n+m$ is an odd integer. Thus, in this
case Proposition 1 remains valid and the recursion operator can be found from
(\ref{e00}) or (\ref{e01}).

On the other hand, if $L^{*}=-L$ then both integers $m$ and $n$
are odd and hence their sum $m+n$ is an even integer. This means
that $(L^{n+m \over m})_{+}$ can not be taken as an $A_{n}$-operator. In this
(skew adjoint) case we must take

$$A_{n+2m}=(L^{n+2m \over m})_{+}=(L^2\,L^{n \over m})_{+},$$

\noindent
to find the recursion operator.
Following the proof of Proposition 1 we obtain

\vspace{0.3cm}

\noindent
{\bf Proposition 2}. \, If $L^{*}=-L$ then

\begin{equation}
A_{n+2m}=L^2\, A_{n}+R_{n}, \label{s00}
\end{equation}

\noindent
where $ord(R_{n}) < 2 \,\,ord\,(L).$ It follows from (\ref{s00}) that

\begin{equation}
L_{t_{n+2m}}=L^2\,L_{t_{n}}+[R_{n},L]. \label{e03}
\end{equation}

\noindent

\vspace{0.3cm}

\noindent
{\bf Remark 2:}\, Instead of (\ref{s00}) we can use the ansatz

\begin{equation}
A_{n+2m}=L\,A_{n}\,L+ \tilde{R_{n}}, \label{s01}
\end{equation}

\noindent
or

\begin{equation}
A_{n+2m}=A_{n}\,L^2+\tilde{\tilde{R_{n}}}. \label{s02}
\end{equation}

\noindent
The recursion operators obtained by the utility of (\ref{s00}),
(\ref{s01}), and (\ref{s02}) all coincide.

In the works  \cite{ds1}, \cite{ds2} more general
reductions $L^{\dagger}=\pm L$ were also considered. Here $L^{\dagger}=
KL^{*}K^{-1}$ where $K$ is a given differential operator, such that $LK^{-1}$ is a differential operator.
In this general reductions, as well, possible $A_{n}$-operators are given
by (\ref{a2}) with $n$ being an odd integer. The Propositions 1 and
2 are valid for this general symmetric and skew-symmetric cases
and hence one can use the equations (\ref{e00}), (\ref{e03}) accordingly
to obtain the recursion operators.

\vspace{0.3cm}

\noindent
{\bf Example 2.\, Kupershmidt equation}. This equation

\begin{equation}
 u_{t}=u_{5x}+10uu_{3x}+25u_{x}u_{2x}+20u^{2}u_{x}
\end{equation}
has the Lax pair
\begin{equation}
L=D^3+2u\,D +u_{x},~~~A=(L^{5 \over 3})_{+}.
\end{equation}

\noindent
In this case  $L^{*}=-L$  therefore we use the equation (\ref{e03}) with

\begin{equation}
\tilde{R}_{n}=a_{n}D^{5}+b_{n}D^{4} +c_{n}D^{3}+d_{n}D^{2}+e_{n}D+f_{n}.
\label{r11}
\end{equation}

\noindent
By equating the coefficients of powers of $D$ in (\ref{e03}) we obtain

\begin{eqnarray}
a_{n}&=&{2 \over 3}D^{-1}(u_{n}),~~b_{n}={11 \over 3}u_{n},
~~c_{n}={1 \over 9}(20uD^{-1}(u_{n}) +73u_{n,x}), \nonumber\\
d_{n}&=&{1 \over 3}(10u_{x}D^{-1}(u_{n}) +22uu_{n}+27u_{n,2x}), \nonumber \\
e_{n}&=&{1 \over 27}(70u_{2x}D^{-1}(u_{n}) -2D^{-1}(u_{2x}u_{n})
      +40u^{2}D^{-1}(u_{n})-8D^{-1}(u^{2}u_{n}) \nonumber \\
 & + & 134u_{n,3x}+212uu_{n,x} +184u_{x}u_{n}, \nonumber \\
f_{n,x}&=&{1 \over 27}(20u_{4x}D^{-1}(u_{n}) +74u_{3x}u_{n}
+126u_{2x}u_{n,x}+40uu_{2x}D^{-1}(u_{n}) \nonumber \\
&+&40u_{x}^{2}D^{-1}(u_{n})
+136u_{x}u_{n,2x}+27uu_{x}u_{n}+28u_{n,5x}+64uu_{n,3x}\nonumber \\
&+&16u^{2}u_{n,x}), \nonumber
\end{eqnarray}

\noindent
and the recursion operator for the Kupershmidt equation:

\begin{eqnarray}
 {\cal R} & = & D^{6}+12uD^{4}+36u_{x}D^{3}+(49u_{2x}+36u^{2})D^{2}
\nonumber \\
&+&5(7u_{3x}+24uu_{x})D
+13u_{4x}+82uu_{2x}+69u_{x}^{2}+32u^{3}+ \nonumber \\
& &2u_{x}D^{-1}(u_{2x}+4u^{2})
+2(u_{5x}+10uu_{3x}+25u_{x}u_{2x}+20u^{2}u_{x})D^{-1}.
\end{eqnarray}

\subsection{Pseudo-differential Lax operator.}

In this section we generalize our scheme
to the case of pseudo-differential Lax operators. The only difference
is that in formulas like (\ref{aa}) and (\ref{s00}) the $R_{n}$ operator
becomes also a pseudo-differential operator.

It follows from these formulas that the structure of the
nonlocal terms in $R_{n}$ is, in general, similar to the nonlocal
terms in $L$ since $A_{n+m}$ and $A_{n}$ are differential operators.

For skew-symmetric case, $A_{n}$ may be defined by either (\ref{s00})
or (\ref{s01}) , or (\ref{s02}).  In the
pseudo-differential case they are not equivalent in the sense that the
nonlocal part of $R_{n}$ depends on which ansatz we choose.
For illustration, let us consider the case $L=M\,D^{-1}$, where $M$
is a differential operator. The following lemma shows that if $L^{\dagger}
=L$ or $L^{\dagger}=-L$, where

\begin{equation}
L^{\dagger}=D\, L^{*}\,D^{-1}, \label{x01}
\end{equation}

\noindent
then the formulas (\ref{aa}) and (\ref{s01}) are much suitable
then (\ref{x00}),(\ref{s00}), and (\ref{s02}).

\vspace{0.3cm}

\noindent
{\bf Lemma}. Let $L^{\dagger}=\varepsilon L$, where $\varepsilon =\pm 1$.
Then

\begin{equation}
R_{n}=D^{m-1}+\cdots+a_{0} ,~~~ for\,\, \varepsilon=1 , \label{r01}
\end{equation}

\noindent
where $R_{n}$ is defined by (\ref{aa}), and

\begin{equation}
\tilde{R}_{n}=D^{2m-1}+\cdots+a_{-1}\,D^{-1} , ~~~for \,\,\varepsilon=-1 ,
\label{r00}
\end{equation}

\noindent
where $\tilde{R}_{n}$ is defined by (\ref{s01}).

\vspace{0.3cm}

\noindent
{\bf Proof}: If $L=M\,D^{-1}$ then $L^{\dagger}= \varepsilon\,L$
implies $M^{*}=-\varepsilon M$. It is easy to show that
$(L^{1 \over m})^{\dagger}=-L^{1 \over m}$. Hence
$(L^{n \over m})^{\dagger}=-L^{n \over m}$  for an odd integer $n$.
Define now a series $K_{n}$ by

$$L^{n \over m}=D\,K_{n}. $$

\noindent
It is easy to prove that $K_{n}^{*}=K_{n}$. Since $K_{n}=(K_{n})_{+}+
(K_{n})_{-}$ and $(K_{n})^{*}=K_{n}$, we have

$$(K_{n})_{+}^{*}=(K_{n})_{+},~~~ (K_{n})_{-}^{*}=(K_{n})_{-}. $$

\noindent
From the last formula it follows that $ord(K_{n})_{-} \le -2$ that
leads to an important result

$$A_{n}=(L^{n \over m})_{+}=D\,(K_{n})_{+}. $$

\noindent
This implies that

\begin{equation}
LA_{n}=M(K_{n})_{+}, \label{s06}
\end{equation}

\noindent
is a differential operator. Now using (\ref{s06}) in
(\ref{aa}) and (\ref{s01})
for the cases $\varepsilon=1$ and  $\varepsilon=-1$ respectively
we find the ansatz for $A_{n}$ given by (\ref{r01}) and  (\ref{r00}).

\vspace{0.3cm}

\noindent
{\bf Example 3}~ ($\varepsilon=-1$). It is known that the KdV equation has,
besides the standard Lax representation, the following Lax pair:

\begin{equation}
L=(D^2+u)\,D^{-1}, ~~~~A=(L^3)_{+}. \label{s03}
\end{equation}

\noindent
The $L$-operator satisfies the reduction $L^{\dagger}=-L$.
According to the formula (\ref{r00}) we have

$$\tilde{R_{n}}=a_{n}\,D+b_{n}+c_{n}D^{-1}. $$

\noindent
It follows from (\ref{s01}) that

$$a_{n}=D^{-1}\,(u_{n}),~~~~b_{n}=u_{n}, ~~~~
c_{n}=-u_{n,x}-u\,D^{-1}\,(u_{n}). $$

\noindent
The remaining equation in (\ref{s01}) gives the recursion operator

\begin{equation}
{\cal{R}}=D^{2}+4u+2u_{x}\,D^{-1}.
\end{equation}

\vspace{0.3cm}

\noindent
{\bf Example 4}~ ($\varepsilon=1$). {\bf DSIII system}. \, The DSIII system
 \cite{ds1} \cite{ds2} is given by

\begin{eqnarray}
u_{t} & = & -u_{3x}+6 uu_{x}+6v_{x} ,\nonumber\\
v_{t} & = & 2 v_{3x}-6uv_{x}.
\end{eqnarray}

\noindent
The nonlocal Lax representation for this system is

\begin{eqnarray}
L&=&(D^5-2u\,D^3-2D^3\,u-2D\,w-2w\,D)D^{-1} ,\\
A&=&(L^{3 \over 4})_{+} \nonumber ,
\end{eqnarray}

\noindent
where $w=v-u_{2x}$. Since $L^{\dagger}=L$  we can use (\ref{r01})
which gives us

\begin{equation}
R_{n}=a_{n}\,D^3+b_{n}D^2+c_{n}D+d_{n}. \label{r14}
\end{equation}

\noindent
By equating the coefficients of the powers of $D$ in (\ref{s01}) we obtain

\begin{eqnarray}
a_{n}&=&D^{-1}\,(u_{n}),~~~b_{n}=4\,u_{n}, \nonumber \\
c_{n}&=&{1 \over 2}(-6\,u\,D^{-1}\,(u_{n})+11\,u_{n,x}+2\,D^{-1}\,(u\,u_{n})
+2\,D^{-1}\,(v_{n})), \nonumber \\
d_{n,x}&=&-{1 \over 2}\,(6\,u_{2x}\,D^{-1}\,(u_{n})+10\,u_{x}\,u_{n}
-5\,u_{n,3x}+4\,u\,u_{n,x}-6\,v_{n,x}). \nonumber
\end{eqnarray}

\noindent
The recursion operator of the DSIII is found as

\begin{equation}
{\cal R}=
\left(
\begin{array}{cc}
{\cal R}^{0}_{0}
 & {\cal R}^{0}_{1} \\
{\cal R}^{1}_{0}  &
{\cal R}^{1}_{1}
\end{array} \; \; \right)\;
  \label{rec}
\end{equation}

\noindent
with

\begin{eqnarray}
{\cal R}^{0}_{0} & = & D^{4}- 8 uD^{2}
-12u_{x}D-8u_{2x}+16u^{2}
+16v+(-2\,u_{3x}+12\,uu_{x}+ \nonumber \\
& & 12\,v_{x})D^{-1}+4u_{x}D^{-1}u,\nonumber\\
{\cal R}^{0}_{1} & = & -10D^{2}+8u+4u_{x}D^{-1},
\nonumber\\
{\cal R}^{1}_{0} & = & 10v_{x}D
+12v_{2x}+(4v_{3x}-12uv_{x})D^{-1}+4v_{x}D^{-1}u,\nonumber\\
{\cal R}^{1}_{1} & = & -4D^{4}+16uD^{2}+8u_{x}D+16v+4v_{x}D^{-1}.
\label{z1}
\end{eqnarray}

\noindent
This recursion operator has recently been given in \cite{kar}.

\section{Matrix L-operator of the first order.}

In this section we demonstrate how our approach, given in
the previous sections, can be generalized to the case where
$L$ is a matrix operator of the form

\begin{equation}
L=D_x+\lambda a+q(x,t). \label{la1}
\end{equation}

\subsection{General Case}

Let us consider the Lax operator  (\ref{la1}),
where $q$ and $a$ belong to a Lie algebra $ {\it g}$ and
$\lambda$ is the spectral parameter. The constant element $a$ is
supposed to be such that

\begin{equation}
{\it g}=Ker \,(ad_{a}) \oplus Im \,(ad_{a}).
\end{equation}

\noindent
First let us recall  the procedure \cite{ds1} of constructing the $A$-
operators for the Lax operator (\ref{la1}).

\vspace{0.3cm}

\noindent
{\bf Proposition 3}. There exist unique series

\begin{eqnarray}
u&=&u_{-1}\, \lambda^{-1}+ u_{-2}\, \lambda^{-2}+\cdots , ~~~~~~~~~~~
u_{i} \epsilon \,Im \,(ad_{a}), \label{la2}  \\
h&=&h_{0}+h_{-1}\, \lambda^{-1}+ h_{-2}\, \lambda^{-2}+\cdots , ~~~
~~h_{i} \epsilon \,Ker \,(ad_{a}),  \label{la3}
\end{eqnarray}

\noindent
such that

\begin{equation}
e^{ad_{u}}\,(L)=L+[u,L]+{1 \over 2}\,[u,[u,L]]+\cdots =D_{x}+a\lambda+h.
\end{equation}

Let $b$ be a constant element of ${\it g}$ such that
$[b, Ker\,(ad_{a})]=\{0\}$. It follows from (\ref{la3}) that
$[b\, \lambda^{n}, D_{x}+a \lambda+h]=0$. Hence $[\Phi_{b,n},L]=0$, where

\begin{equation}
\Phi_{b,n}= e^{-ad_{u}}\,(b\, \lambda^{n}). \label{la4}
\end{equation}

\noindent
Then the corresponding  $A$-operator of the form

\begin{equation}
A_{b,n}=b\, \lambda^{n}+a_{n-1}\, \lambda^{n-1}+\cdots+a_{0},
\end{equation}

\noindent
is defined by the formula

\begin{equation}
A_{b,n}=(\Phi_{b,n})_{+}, \label{la5}
\end{equation}

\noindent
where

\begin{equation}
(\Sigma_{-\infty}^{n}\, \alpha_{i}\, \lambda^{i})_{+}=
\Sigma_{0}^{n}\, \alpha_{i}\, \lambda^{i}.  \label{ser1}
\end{equation}

\noindent
According to  (\ref{la4})

\begin{equation}
\Phi_{b,n+1}=\lambda\, \Phi_{b,n}. \nonumber
\end{equation}

\noindent
Hence

\begin{equation}
A_{b,n+1}=(\lambda\, \Phi_{b,n})_{+}=\lambda\, (\Phi_{b,n})_{+}+
(\lambda\, (\Phi_{b,n})_{-})_{+}. \nonumber
\end{equation}

\noindent
The last formula shows that

\begin{equation}
A_{b,n+1}=\lambda\, A_{b,n}+R_{n},~~~~~ R_{n} \in {\it g}, \label{y00}
\end{equation}

\noindent
where $R_{n}$ does not depend on $\lambda$. Substituting (\ref{y00})
into the Lax equation $L_{t_{n+1}}=[A_{b,n+1},L]$ we get

\begin{equation}
L_{t_{n+1}}=\lambda \, L_{t_{n}}+ [R_{n},L]. \label{y01}
\end{equation}

\noindent
Using the ansatz (\ref{y01}) one can easily find the corresponding
recursion operator.

\vspace{0.3cm}

\noindent
{\bf Example 5.}\, The system

\begin{eqnarray}
u_{t}&=&-{1 \over 2}\,u_{xx}+u^2\,v, \nonumber\\
v_{t}&=&{1 \over 2}\,v_{xx}-v^2\,u  , \label{yy0}
\end{eqnarray}

\noindent
is equivalent to the nonlinear Schr{\" o}dinger equation, has
a Lax operator

\begin{equation}
L=D+\left( \matrix{ 1 & 0 \cr
                        0 & -1 \cr} \right) \, \lambda +
          \left( \matrix{ 0 & u \cr
                       v & 0 \cr } \right) . \label{y02}
\end{equation}

\noindent
The Lie algebra ${\it g}$ in this example coincides with $sl(2)$.

\noindent
Using (\ref{y01}) with $ R_{n}=\left( \matrix{a_{n} & b_{n} \cr
               c_{n} & -a_{n} \cr} \right) $
we find that

\begin{eqnarray}
a_{n}&=&{ 1 \over 2}\,D^{-1}\,(v u_{n}+u v_{n}), \nonumber \\
b_{n}&=&{1 \over 2}\,u_{n},~~~~c_{n}=-{1 \over 2}\,v_{n}, \nonumber
\end{eqnarray}

\noindent
and the recursion operator of the system (\ref{yy0}) is given by

\begin{equation}
\cal{R}=\left( \matrix{-{1 \over 2}\,D+u\, D^{-1}v & u\,D^{-1}u \cr
 -v\,D^{-1} v & {1 \over 2}\, D-v\,D^{-1}u \cr} \right). \label{y03}
\end{equation}

\subsection{Reductions in matrix case}

In the general case considered in the previous section the $A_{n}$-operators
belong to Lie algebra

\begin{equation}
{\frak{a}}_{+}=\{\Sigma_{i=0}^{\kappa}\,a_{i}\, \lambda^{i},~~a_{i} \in
{\it g},~~ \kappa \in Z_{+}\}, \label{al1}
\end{equation}

\noindent
that is a sub-algebra of the Lie algebra

\begin{equation}
{\frak{a}}=\{\Sigma_{-\infty}^{\kappa}\,a_{i}\, \lambda^{i},~~a_{i} \in
{\it g},~~ \kappa \in Z \}. \label{al2}
\end{equation}

A standard $\sigma$-reduction is defined by any automorphism $\sigma$
of the Lie algebra ${\it g}$ of finite order $\kappa$. Because
$\sigma^{\kappa}=Id$, the eigenvalues of $\sigma$ are $\varepsilon ^{i},
i=0, \cdots ,\kappa-1$ where $\varepsilon$ is primitive $\kappa$-root of unity.

Let ${\it g}_{i}$ be a eigenspace corresponding to eigenvalue
$\varepsilon^{i}$. Then the following reduction  $a_{j} \in {\it g}_{i}$,
where $i=j\, (mod \, \kappa)$ in (\ref{al1}) and (\ref{al2})
is compatible with the equations (\ref{l2}). Note that according to this
definition  $a \in {\it g}_{1}$  and the potential $q(x,t)$ in (\ref{la1}) belongs to ${\it g}_{0}$ or,
the same, satisfies $\sigma(q)=q$.

It is easy to see that, to satisfy such a reduction, we must use the ansatz

\begin{equation}
A_{b,n+\kappa}=\lambda^{\kappa}\, A_{b,n}+R_{n}, \label{al7}
\end{equation}

\noindent
where

\begin{equation}
R_{n}=r_{\kappa-1}\, \lambda^{\kappa-1}+\cdots +r_{0},~~~
r_{i} \in {\it g}_{i}. \label{al3}
\end{equation}

Further generalizations are associated with  modifications of sign
$"+"$ in (\ref{ser1}) which corresponds to the
simplest decomposition of algebra $\frak{a}$ into direct sum of two
sub-algebras

\begin{equation}
{\frak{a}}={\frak{a}}_{+} \oplus {\frak{a}}_{-}, \label{al4}
\end{equation}

\noindent
where $\frak{a}_{+}$ is given by (\ref{al1}) and

\begin{equation}
{\frak{a}}_{-}=\{\Sigma_{-\infty}^{-1}\,a_{i}\, \lambda^{i},
~~a_{i} \in {\it g}\}. \label{al5}
\end{equation}

\noindent
The sign $"+"$ in (\ref{ser1}) is the projection of onto $\frak{a}_{+}$
parallel to $\frak{a}_{-}$. If we have a different decomposition (\ref{al4})
then the construction from Proposition 3  is also valid but
we have the following condition

\begin{equation}
R_{n} \in \frak{a}_{+} \cap  \lambda\, \frak{a}_{-}, \nonumber
\end{equation}

\noindent
instead of $R_{n} \in {\it g}$. If we also have the $\sigma$-reduction
we must use the most general ansatz (\ref{al7}) where

\begin{equation}
R_{n} \in \frak{a}_{+} \cap  \lambda^{\kappa}\, \frak{a}_{-}. \label{al6}
\end{equation}

\vspace{0.3cm}

\noindent
{\bf Example 6}. Let us consider the following equation

\begin{equation}
u_{t}={1 \over 4}\,u_{xxx}-{3 \over 8}\,u_{xx}\,u+{3 \over 8}\, u\,u_{xx}
-{3 \over 8}\, u\,u_{x}\,u , \label{or1}
\end{equation}

\noindent
where $u$ is a square matrix of arbitrary size, or more generally, $u$ belongs
to arbitrary associative algebra ${\cal K}$. This equation has a Lax
representation with

\begin{equation}
L=D+\left(\matrix{0 & \bf{1} \cr
                         \bf{1} & 0 \cr} \right)\, \lambda+
                         \left( \matrix{ u & 0  \cr
                                       0 & 0 \cr} \right).  \label{or2}
\end{equation}

\noindent
Here $\bf{1}$ is the unity of ${\cal{K}}$. The reduction (\ref{or2})
can be described as follows (see \cite{sokg}). The Lie
algebra ${\it g}$ is the algebra of all
$2 \times 2$ matrices with entries belonging to ${\cal{K}}$. The
automorphism $\sigma$ is defined by

\begin{equation}
\sigma(X)=T\,X\,T^{-1},
\end{equation}

\noindent
where
$T=\left(\matrix{\bf{1} & 0 \cr
                 0 & -\bf{1} \cr} \right)$.
Obviously $\sigma^2=Id$ and eigenvalues of $\sigma$ are $1$ and $-1$.
The corresponding eigenspaces are

\begin{equation}
{\it g}_{0}=\left \{ \matrix{ * & 0 \cr
                      0 & * \cr} \right \}, ~~~~~
{\it g}_{1}=\left \{ \matrix{ 0 & * \cr
                      * & 0 \cr} \right \}, \label{or3}
\end{equation}

\noindent
and therefore  the coefficients $a_{i}$ in (\ref{al2}) have the following
structure

\begin{equation}
a_{2j}=\left(\matrix{* & 0 \cr
                0 & * \cr} \right), ~~~~~
a_{2j+1}=\left(\matrix{0 & * \cr
                * & 0 \cr} \right). \label{or4}
\end{equation}

\noindent
The sub-algebra ${\frak{a}}_{+}$ is given by  (\ref{al1}), where
the coefficients have the structure (\ref{or4}) and, additionally
$a_{0}=\left(\matrix{* & 0\cr
                     0 & 0 \cr} \right)$.
The sub-algebra $\frak{a}_{-}$ has the following form

\begin{equation}
{\frak{a}}_{-}= \Sigma_{-\infty}^{0}\, a_{i}\, \lambda^{i},
\end{equation}

\noindent
where $a_{0}$ is of the form
$a_{0}=\left(\matrix{\alpha & 0\cr
                     0 & \alpha \cr} \right), ~~~\alpha \in \cal{K}$.

The $A$-operator for (\ref{or1}) is given by formula
$A=(\Phi_{a,3})_{+}$ (see (\ref{la5})), where
$a=\left( \matrix{0 & \bf{1} \cr
                 \bf{1} & 0 \cr} \right)$
and $"+"$ means the projection onto ${\frak{a}}_{+}$ parallel to
${\frak{a}}_{-}$.

According to (\ref{al6}), $R_{n}$ is of the form

\begin{equation}
R_{n}=\left( \matrix{ a_{n} & 0 \cr
                     0 & a_{n} \cr} \right)\,\lambda^2+
\left( \matrix{ 0 & b_{n} \cr
                     c_{n} & 0 \cr} \right)\, \lambda+
\left( \matrix{ d_{n} & 0 \cr
                     0 & 0 \cr} \right). \nonumber
\end{equation}

\noindent
It follows from

\begin{equation}
L_{t_{n+2}}=\lambda^2\, L_{t_{n}}+[R_{n},L],
\end{equation}

\noindent
that

\begin{eqnarray}
u_{n}-a_{n,x}+[a_{n},u]+b_{n}-c_{n}=0, ~~~~c_{n}-b_{n}-a_{n,x}=0, \nonumber\\
d_{n}-b_{n,x}-u\,b_{n}=0,~~~~d_{n}+c_{n,x}-c_{n}\,u=0, \nonumber \\
u_{n+2}=-d_{n,x}+[d_{n},u]. \nonumber
\end{eqnarray}

\noindent
Finding $a_{n}, b_{n}, c_{n}$, and $d_{n}$ from this system
we obtain the following recursion operator

\begin{equation}
{\cal{R}}=-(D+ad_{u})\,(-D+R_{u})\,(2\,D+ad_{u})^{-1}\, (D
+L_{u})\, D\, (2\,D+ad_{u})^{-1}, \label{or5}
\end{equation}

\noindent
where $R_{u}$ and $L_{u}$ are the operators of right and
left multiplications by $u$, respectively.

Note that in the commutative case (\ref{or1}) coincides with
the modified KdV equation. It is easy to verify that (\ref{or5})
becomes the standard recursion operator of modified KdV equation.
All factors in (\ref{or5}) have to be regarded as operators
acting on (non-commutative) polynomial depending on $u,u_{x},
u_{xx}, \cdots$.

\section{Conclusion}

In this work we devoted ourselves in the construction of
recursion operators when the Lax representation is given.
We have shown that our approach can be easily generalized
to all cases where $L$-operator is a polynomial of $\lambda$.
It would be interesting to generalize it for the cases of more complicated
$\lambda$ dependence of $L$ as well as for the cases $2+1$ dimensional
equations, Toda-type lattices and ordinary differential equations.

\section*{Acknowledgments}

We would like to thank Dr. Jing Ping Wang for reading the manuscript
and pointing out some misprints.
This work is partially supported by the Scientific and Technical
Research Council of Turkey (TUBITAK) and Turkish Academy of Sciences 
(TUBA). V.S is supported by RFFR - Grant 99-01-00294 and INTAS.

\section{APPENDIX A}

In this Appendix we give an additional example to Section 2.1.

\subsection*{A. Boussinesq System.}

The Boussinesq equation

\begin{equation}
u_{tt}=-{1 \over 3}(u_{4x}+2(u^{2})_{2x})
\end{equation}

\noindent
can be expressed in the form of a pair of first-order evolution equations

\begin{eqnarray}
u_{t} & = &v_{x} ,\nonumber\\
v_{t} & = &- {1 \over 3}(u_{3x}+8uu_{x}). \label{bus}
\end{eqnarray}

\noindent
This system has  a Lax pair

\begin{equation}
L=D^{3}+2uD+u_{x}+v, ~~~A=(L^{2 \over 3})_{+}.
\end{equation}

\noindent
To construct the recursion operator for this system we use the
 equation (\ref{e00}) with the differential operator

$$R_{n}=a_{n}D^{2}+b_{n}D +c_{n}. $$

\noindent
By equating the coefficients of the powers of $D$ in (\ref{e00}) we find

\begin{eqnarray}
a_{n}={2 \over 3}D^{-1}(u_{n}),~~~b_{n}={1 \over 3}(5u_{n}
+D^{-1}(v_{n})), \nonumber \\
c_{n}={1 \over 9}(6v_{n}+8uD^{-1}(u_{n}) +10u_{n,x}), \nonumber
\end{eqnarray}

\noindent
and after that we obtain the recursion operator of the form
(\ref{rec}) for (\ref{bus}) with

\begin{eqnarray}
{\cal R}^{0}_{0} & = &3v+2 v_{x}\,D^{-1}, \nonumber\\
{\cal R}^{0}_{1} & = & D^{2}+2u+u_{x}D^{-1}, \nonumber\\
{\cal R}^{1}_{0} & = &-( {1 \over 3}D^{4}+{10  \over3}uD^{2}
+5u_{x}D+3u_{2x}+{16 \over 3}u^{2}+({2 \over 3}u_{3x}+{16 \over 3}uu_{x}
)D^{-1}),\nonumber\\
{\cal R}^{1}_{1} & = &3v+v_{x}D^{-1}.  \label{a18}
\end{eqnarray}

\section{APPENDIX B}

In this Appendix we give additional examples to the Section 2.2.

\subsection*{B.1. Sawada-Kotera Equation.}

The Lax pair for the Sawada-Kotera equation  \cite{sk}

\begin{equation}
u_{t}=u_{5x}+5uu_{3x}+5u_{x}u_{2x}+5u^2u_{x},
\end{equation}

\noindent
is given by

\begin{equation}
L=D^3+uD, ~~~A=(L^{5 \over 3})_{+}.
\end{equation}

\noindent
In this example, $L^{\dagger}=-L$, where $L^{\dagger}=D^{-1} L^{*}\, D$
and $L$ is skew-symmetric, then we use (\ref{e03}).
The operator $\tilde{R}_{n}$ has the same form as (\ref{r11}) with
the coefficients given by

\begin{eqnarray}
a_{n}&=&{1 \over 3}\, D^{-1}\,(u_{n}),~~b_{n}= {5 \over 3} u_{n},
~~~c_{n}={1 \over 9}\, (5uD^{-1}(u_{n})+29u_{n,x}), \nonumber \\
d_{n}&=&{1 \over 9}\,(5u_{x}\,D^{-1}(u_{n})+14uu_{n}+26u_{n,2x}), \nonumber \\
e_{n}&=&{1 \over 27}\,(10u_{2x}\, D^{-1}(u_{n})-2 D^{-1}\,(u_{2x}u_{n})-
D^{-1}(u^2 u_{n})+5u^2 D^{-1}(u_{n}) \nonumber \\
&+&28 u_{n,3x}+32u u_{n,x}-32u_{x}u_{n}),~~~f_{n}=0. \nonumber
\end{eqnarray}

\noindent
The recursion operator is given as

\begin{eqnarray}
{\cal{R}}=D^6+6u D^4+9u_{x}D^3+(9u^2+11u_{2x})D^2+(10u_{3x}+21uu_{x})D \nonumber \\
+5u_{4x}+16uu_{2x}+6u_{x}^2+4u^3+(u_{5x}+5uu_{3x}+5u_{x}u_{2x}+ \nonumber \\
5u^2u_{x})D^{-1}+u_{x}D^{-1} (u^{2}+2u_{2x}).
\end{eqnarray}

\subsection*{B.2. DSI System.}

The DSI system \cite{ds1},\cite{ds2}

\begin{eqnarray}
u_{t}&=&3vv_{x}, \nonumber \\
v_{t}&=&2v_{3x}+2uv_{x}+vu_{x}, \label{a02}
\end{eqnarray}

\noindent
has a Lax representation with

\begin{eqnarray}
L=[D^3+(u+v)\,D+{1 \over 2}\,(u+v)_{x}]\,[D^3+ (u-v)\,D+
{1 \over 2}\,(u-v)_{x}],\\
A=(L^{1 \over 2})_{+}. \nonumber
\end{eqnarray}

\noindent
Here $R_{n}$ is a differential operator of order $5$ and since $L$ is
symmetric we again use the equation (\ref{e00}) . The expressions for
the coefficients of the operator $R_{n}$ are very long and complicated.
Hence we do not display them here.

\noindent
We find that the recursion operator ${\cal R}$ of this system is
of the form (\ref{rec})

\noindent
where

\begin{eqnarray}
{\cal R}^{0}_{0}&=& -4D^{6}-24uD^{4}-72u_{x}D^{3}
+2(-49u_{2x}-18u^{2}+42v^{2})D^{2} \nonumber\\
&+&10(-7u_{3x}-12uu_{x}+30vv_{x})D \nonumber \\
&-&26u_{4x}-82uu_{2x}
-69u_{x}^{2}+222vv_{x}+141v_{x}^{2}-16u^{3}+48v^{2}u \nonumber \\
&+&2(-2u_{5x}-10uu_{3x}-25u_{x}u_{2x}-10u^{2}u_{x}+15v^{2}u_{x}
+30vv_{3x} \nonumber \\
&+&45v_{x}v_{2x}+30uvv_{x})D^{-1}+ 2 u_{x}D^{-1}(3v^{2}-2u^{2}-u_{2x}),
 \nonumber \\
{\cal R}^{0}_{1} & = & 168vD^{4}+204v_{x}D^{3}+6(21v_{2x}+32uv)D^{2}
+6(40vu_{x}+7v_{3x}+22uv_{x})D \nonumber \\
&+&6(13vu_{2x}+10u_{x}v_{x}+v_{4x}+5uv_{2x}+4vu^{2}+12v^{3}) \nonumber
\\
&+& 108vv_{x}D^{-1}v+2u_{x}D^{-1}(6uv+9v_{2x}),
\nonumber\\
{\cal R}^{1}_{0} & = & 56vD^{4}+268v_{x}D^{3}+2(243v_{2x}+32uv)D^{2}
 \nonumber \\
&+&2(36vu_{x}+219v_{3x}+106uv_{x})D+2(27vu_{2x}+92u_{x} v_{x}+99v_{4x}
+99uv_{2x} \nonumber \\
&+& 4vu^{2}+12v^{3})
+2(10vu_{3x}+35u_{2x}v_{x}+45u_{x}v_{2x}+10uvu_{x}+18v_{5x}
\nonumber \\
&+&30uv_{3x}+10u^{2}v_{x}+15v^{2}v_{x})D^{-1}+2v_{x}D^{-1}(3v^{2}-
2u^{2}-u_{2x}),
\nonumber\\
{\cal R}^{1}_{1} & = & 108D^{6}+216uD^{4}+432u_{x}D^{3}
+6(81u_{2x}+18u^{2}+22v^{2})D^{2}\nonumber\\
&+&6(45u_{3x}+36uu_{x}+70vv_{x})D \nonumber \\
&+&3(18u_{4x}+18uu_{2x}+9u_{x}^{2}+98vv_{2x}+67v_{x}^{2}+32uv^{2})
\nonumber\\
&+&36( 2v_{3x}+2v_{x}u+vu_{x})D^{-1}\,v+2v_{x}D^{-1}(6uv+9v_{2x}).
\end{eqnarray}

\subsection*{B.3. DSII System.}

The DSII system  \,\, \cite{ds1}, \cite{ds2}

\begin{eqnarray}
u_{t} & = & 3v_{x} ,\nonumber\\
v_{t} & = &-2(v_{3x}+uv_{x}+vu_{x}),
\end{eqnarray}

\noindent
has a Lax representation with

\begin{eqnarray}
L&=&(D^{6}+uD^{3}+D^{3}u+(v+{1 \over 2}u^{2})D+D(v+{1 \over 2}u^{2}), \\
A&=&(L^{1 \over 2})_{+}. \nonumber
\end{eqnarray}

\noindent
Since $L$ is
symmetric we again use the equation (\ref{e00}).
In this case the operator $R_{n}$ is given as follows

\begin{equation}
R_{n}=a_{n}D^5+b_{n}D^4+c_{n}D^3+d_{n}D^2+e_{n}D,
\end{equation}

\noindent
where

\begin{eqnarray}
a_{n}&=&{1 \over 3}\,D^{-1}(u_{n}),~~~b_{n}={5 \over 3}\,u_{n}, \nonumber\\
c_{n}&=&{1 \over 9}[5uD^{-1}(u_{n})+3D^{-1}(v_{n})+29u_{n,x}],\nonumber \\
d_{n}&=&{1 \over 9}[5u_{x}\,D^{-1}(u_{n})+26u_{n,2x}+14uu_{n}+12v_{n}], \nonumber \\
e_{n}&=&{1 \over 27}[5(2u_{2x}+u^2+3v)D^{-1}(u_{n})-3D^{-1}(vu_{n}+uv_{n})+
9uD^{-1}(v_{n}) \nonumber \\
&-&2D^{-1}(u_{2x}u_{n}+{1 \over 2}u^2u_{n})
+54u_{x}u_{n}+28u_{n,3x}+32(uu_{n,x}-u_{n}u_{x})+42v_{n,x}]. \label{r13}
 \nonumber
\end{eqnarray}

\noindent
The recursion operator (\ref{rec}) for the system can be found as

\begin{eqnarray}
 {\cal R}^{0}_{0} & = & -D^{6}-6uD^{4}-9u_{x}D^{3}
-(11u_{2x}+9u^{2}+42v)D^{2}
\nonumber\\
&+&(-10u_{3x}-21uu_{x}-84v_{x})D-5u_{4x}-16uu_{2x}
-6u_{x}^{2}-60v_{2x}-4u^{3}-24vu \nonumber \\
&+&(-u_{5x}-5uu_{3x}-5u_{x}u_{2x}-5u^{2}u_{x}-15vu_{x}
-15v_{3x}-15uv_{x})D^{-1} \nonumber \\
&-& u_{x}D^{-1}(2u_{2x}+u^{2}+3v), \nonumber \\
 {\cal R}^{0}_{1} & = & -42D^{4}-48uD^{2}-87u_{x}D-6(7u_{2x}+u^{2}-6v)
+27v_{x}D^{-1}-3u_{x}D^{-1}u, \nonumber \\
{\cal R}^{1}_{0} & = & 28vD^{4}+106v_{x}D^{3}+(165v_{2x}+32uv)D^{2}
\nonumber \\
&+&(54vu_{x}+132v_{3x}+74v_{x}u)D \nonumber \\
&+& 30vu_{2x}+79u_{x}v_{x}+54v_{4x}+57uv_{2x}+4u^{2}v-24v^{2}\nonumber\\
&+&(10vu_{3x}+25v_{x}u_{2x}+30u_{x}v_{2x}+10uvu_{x}+9v_{5x}+15uv_{3x} \nonumber \\
&+&5u^{2}v_{x}-15vv_{x})D^{-1}-v_{x}D^{-1}(3v+u^{2}+2u_{2x}), \nonumber\\
{\cal R}^{1}_{1} & = & 27D^{6}+54uD^{4}+135u_{x}D^{3}
+3(54u_{2x}+9u^{2}-22v)D^{2}
 \nonumber \\
&+&3(36u_{3x}+27uu_{x}-28v_{x})D+3(9u_{4x}+9uu_{2x} +9u_{x}^{2}
-21v_{2x}-16vu) \nonumber \\
&-&18(v_{3x}+u_{x}v+v_{x}u)D^{-1}-3v_{x}D^{-1}u .
\end{eqnarray}

\subsection*{B.4. DSIV System.}

The DSIV System \, \cite{ds1}, \cite{ds2} \,which is also known as
the Hirota-Satsuma system \cite{hs0},\cite{wil}

\begin{eqnarray}
u_{t} & = & {1 \over 2}u_{3x}+3uu_{x}-6vv_{x} ,\nonumber\\
v_{t} & = &-v_{3x}-3uv_{x}, \label{a12}
\end{eqnarray}

\noindent
has Lax representation with

\begin{equation}
L=(D^{2}+u+v)(D^{2}+u-v)~~~,~~A=(L^{3 \over 4})_{+}.
\end{equation}

\noindent
Since the operator $L$ is symmetric we use the equation (\ref{e00}).
In this case the operator $R_{n}$ has the same form as (\ref{r14})
with coefficients given by

\begin{eqnarray}
a_{n}&=&{1 \over 2}\,D^{-1}(u_{n})~~~,~~b_{n}=
{7 \over 4}\,u_{n}-{1 \over 2}\,v_{n}, \nonumber\\
c_{n}&=&{1 \over 8}[6uD^{-1}(u_{n})+2D^{-1}(uu_{n})-4D^{-1}(vv_{n})+
17u_{n,x}-12v_{n,x}],\nonumber \\
d_{n,x}&=&{1 \over 16}[6u_{2x}\,D^{-1}(u_{n})-12v_{2x}D^{-1}(u_{n})+
30u_{x}u_{n}-8u_{x}v_{n} \nonumber \\
&+&24uu_{n,x}+15u_{n,3x}-12v_{x}v_{n}-8uv_{n,x}
-20vv_{n,x}-28v_{n,3x}]. \nonumber
\end{eqnarray}

\noindent
The recursion operator (\ref{rec}) for the given system is

\begin{eqnarray}
 {\cal R}^{0}_{0} & = & {1 \over 4}D^{4}+2uD^{2}+3u_{x}D+2u_{2x}+4(u^{2}
-v^{2})\nonumber\\
& & +(3uu_{x}- 6vv_{x}+{1 \over 2}u_{3x})D^{-1}+u_{x}D^{-1}u,
\nonumber\\
{\cal R}^{0}_{1} & = & -5vD^{2}-4v_{x}D-v_{2x}-4uv -2u_{x}D^{-1}v,
\nonumber\\
{\cal R}^{1}_{0} & = & - {5 \over 2}v_{x}D
- 3v_{2x}-(v_{3x}+3uv_{x})D^{-1}+v_{x}D^{-1}u,\nonumber\\
{\cal R}^{1}_{1} & = & -D^{4}-4uD^{2}-2u_{x}D-4v^2-2v_{x}D^{-1}v.
 \label{a13}
\end{eqnarray}

\subsection*{B.5.\, N=3 Hirota-Satsuma System.}

This system is given by \, \cite{hs0}

\begin{eqnarray}
u_{t}&=&{1 \over 4}\, u_{3x}+3uu_{x}+3(-v^2+w)_{x},\nonumber \\
v_{t}&=&-{1 \over 2}\,v_{3x}-3uv_{x},\nonumber \\
w_{t}&=&-{1 \over 2}\,w_{3x}-3uw_{x}.
\end{eqnarray}

\noindent
This is an example for $N=3$ system which covers some other
$N=2$ systems as special cases. For instance letting $w=0$ we get DSIV and
$v=0$ we get DSIII systems.

The corresponding Lax pair is

\begin{equation}
L= (D^2+2u-2v)(D^2+2u+2v)+4w ,~~ A=(L^{3 \over 4})_{+}.
\end{equation}

\noindent
In this case the operator $L$ is symmetric and hence
$R_{n}$ has the same form as (\ref{r14}) with the coefficients

\begin{eqnarray}
a_{n}&=&D^{-1}(u_{n}),~~~b_{n}={7 \over 2}u_{n}+v_{n}, \nonumber \\
c_{n}&=&{1 \over 4}[12uD^{-1}(u_{n})+4D^{-1}(uu_{n}+w_{n}-2vv_{n})+17u_{n,x}
+12v_{n,x}], \nonumber \\
d_{n,x}&=&{1 \over 8}[12u_{2x}D^{-1}(u_{n})+24v_{2x}D^{-1}(u_{n})
+60u_{x}u_{n}+16u_{x}v_{n}+15u_{n,3x} \nonumber \\
&+&48uu_{n,x}+24v_{x}u_{n}
-40v_{x}v_{n}+20v_{n,3x}+16vv_{n,x}+20w_{n,x}]. \nonumber
\end{eqnarray}

\noindent
The recursion operator is given by

\begin{equation}
{\cal R}=
\: \left(
\begin{array}{ccc}
{\cal R}^{0}_{0}
 & {\cal R}^{0}_{1}&  {\cal R}^{0}_{2}\\
{\cal R}^{1}_{0}  &
{\cal R}^{1}_{1}& {\cal R}^{1}_{2}\\
{\cal R}^{2}_{0}&{\cal R}^{2}_{1}&{\cal R}^{2}_{2}
\end{array} \; \; \right)\;,
\label{rec1}
\end{equation}

\noindent
where

\begin{eqnarray}
{\cal R}^{0}_{0}&=&{1 \over 4}D^4+4uD^2+6u_{x}D+4(u_{2x}+4u^2-4v^2
+4w) \nonumber \\
&+&4({1 \over 4}u_{3x}+3uu_{x}-6vv_{x}+3w_{x})D^{-1}+4u_{x}D^{-1}\,u, \nonumber\\
{\cal R}^{0}_{1}&=&-2(5vD^2+4v_{x}D+v_{2x}+8uv+4u_{x}D^{-1}\,v), \nonumber\\
{\cal R}^{0}_{2}&=&5D^2+8u+4u_{x}D^{-1}, \nonumber\\
{\cal R}^{1}_{0}&=&-5v_{x}D-6v_{2x}-2(v_{3x}+6v_{x}u)D^{-1}+4v_{x}D^{-1}\,u, \nonumber\\
{\cal R}^{1}_{1}&=&-D^4-8uD^2-4u_{x}D+8(8w-2v^2)-8v_{x}D^{-1}\,v-8D^{-1}\,w_{x}, \nonumber\\
{\cal R}^{1}_{2}&=&4(v_{x}D^{-1}+2D^{-1}\,v_{x}),\nonumber\\
{\cal R}^{2}_{0}&=&-5w_{x}D-6w_{2x}-2(v_{3x}+6w_{x}u)D^{-1}+4w_{x}D^{-1}\,u, \nonumber\\
{\cal R}^{2}_{1}&=&-16vD^{-1}\,w_{x}-8w_{x}D^{-1}\,v, \nonumber\\
{\cal R}^{2}_{2}&=&-D^4-8uD^2-4u_{x}D+16(w-v^2)+4w_{x}D^{-1}+16vD^{-1}\,v_{x}.
\end{eqnarray}

\section*{APPENDIX C}

In this Appendix we give an additional example to Section 3.

\subsection*{C.1. Non-Abelian Schr{\" o}dinger equation.}

This is the system given by

\begin{eqnarray}
u_{t}&=&-{1 \over 2}\,u_{xx}+u\,v\,u , \nonumber\\
v_{t}&=&{1 \over 2}\,v_{xx}-v\,u\,v  , \label{yy5}
\end{eqnarray}

\noindent
where $u$ and $v$ belong to ${\cal{K}}$ (see Example 6 for the notations)).
The Lax operator of (\ref{yy5})  is given by

\begin{equation}
L=D+\left( \matrix{{\bf 1} & 0 \cr
                        0 & -{\bf 1} \cr} \right) \, \lambda +
          \left( \matrix{ 0 & u \cr
                        v & 0 \cr } \right) . \label{yy6}
\end{equation}

\noindent
The corresponding formula (\ref{y01}) reduces to

\begin{equation}
  \left( \matrix{ 0 & u_{n+1} \cr
                        v_{n+1} & 0 \cr } \right) =
\lambda\,  \left( \matrix{ 0 & u_{n} \cr
                        v_{n} & 0 \cr } \right)+[R_{n},L], \label{yy7}
\end{equation}

\noindent
where

\begin{equation}
R_{n}=     \left( \matrix{a_{n} & b_{n} \cr
                        c_{n} & -a_{n} \cr } \right). \nonumber
\end{equation}

\noindent
The formula (\ref{yy7}) gives us both $a_{n}, b_{n}, c_{n}$
and the recursion operator $\cal{R}$. They are given by

\begin{equation}
a_{n}={1 \over 2}\,D^{-1}\,(u_{n}\,v+u\,v_{n}), ~~~
b_{n}={1 \over 2}\, u_{n}, ~~~c_{n}=-{1 \over 2}\, u_{n}, \nonumber
\end{equation}

\begin{equation}
{\cal{R}}={1 \over 2}\, \left( \matrix{-D+R_{u}\,D^{-1}\,R_{v}+
L_{u}\,D^{-1}\,L_{v}
 & R_{u}\,D^{-1}\,L_{u}+L_{u}\,D^{-1}\,R_{u} \cr
-L_{v}\,D^{-1}\,R_{v}-R_{v}\,D^{-1}\,L_{v} & D-R_{v}\,D^{-1}\,R_{u}
-L_{v}\,D^{-1}\,L_{u} \cr } \right).
\end{equation}

\subsection*{C.2. Non-Abelian Modified KdV Equation.}

The standard non-Abelian Modified KdV equation is given by

\begin{equation}
u_{t}={1 \over 4}\, u_{xxx}- {3 \over 4}\,u_{x}u^2-
{3 \over 4}\, u^2\, u_{x}.
\end{equation}

\noindent
The Lax representation of this equation is given

\begin{equation}
L=D+\left( \matrix{0 & {\bf 1} \cr
                        {\bf 1} & 0 \cr} \right) \, \lambda +
          \left( \matrix{ u & 0 \cr
                        0 & -u \cr } \right) . \label{yy8}
\end{equation}

\noindent
The recursion operator $\cal{R}$ can be found from
(\ref{al7}) and (\ref{al3}). In our case the automorphism $\sigma$
is the same as in Example 6 and formulas (\ref{al7}) and (\ref{al3})
give us

\begin{equation}
  \left( \matrix{ 0 & u_{n+1} \cr
                        v_{n+1} & 0 \cr } \right) =
\lambda^2 \,  \left( \matrix{ 0 & u_{n} \cr
                        v_{n} & 0 \cr } \right)+[R_{n},L], \label{yy9}
\end{equation}

\noindent
where

\begin{equation}
R_{n}=     \left( \matrix{0 & a_{n} \cr
                        b_{n} & 0 \cr } \right)\, \lambda+
                            \left( \matrix{c_{n} & 0 \cr
                        0 & d_{n} \cr } \right).
\end{equation}

\noindent
Using (\ref{yy9}) we find $a_{n},b_{n},c_{n},d_{n}$ from the following:

\begin{eqnarray}
b_{n}-a_{n}=u_{n},~~~~~-a_{n,x}-a_{n}\,u-u\,a_{n}+c_{n}-d_{n}=0, \nonumber \\
-b_{n,x}+b_{n}\,u+u\,b_{n}+d_{n}-c_{n}=0, ~~~~~d_{n,x}+c_{n,x}=
[c_{n}-d_{n}, u],\nonumber\\
u_{n+1}=d_{n,x}+[d_{n},u]. \nonumber
\end{eqnarray}

\noindent
The resulting recursion operator is given by

\begin{equation}
{\cal{R}}={1 \over 4}\, (D -ad_{u} \cdot D^{-1} \cdot ad_{u})\,
(D-(L_{u}+R_{u})\,D^{-1}\, (L_{u}+R_{u})).
\end{equation}

\end{document}